\documentclass[conference]{IEEEtran}
\IEEEoverridecommandlockouts

\usepackage{cite}
\usepackage{amsmath,amssymb,amsfonts}
\usepackage{algorithmic}
\usepackage{graphicx}
\usepackage{textcomp}
\usepackage{xcolor}
\usepackage{multirow}
\usepackage{tabularray}
\usepackage{makecell}
\usepackage{enumitem}
\usepackage{subcaption}

\def\BibTeX{{\rm B\kern-.05em{\sc i\kern-.025em b}\kern-.08em
    T\kern-.1667em\lower.7ex\hbox{E}\kern-.125emX}}
\begin{document}

\title{Managing Risks from Large Digital Loads\\ Using Coordinated Grid-Forming Storage Network
\thanks{Pacific Northwest National Laboratory (PNNL) is operated by Battelle for the DOE under Contract DOE-AC05-76RL01830. This work was supported by PNNL's Lab Directed Research and Development (LDRD) program.}
}

\author{\IEEEauthorblockN{Soumya Kundu, Kaustav Chatterjee, Ramij R.\,Hossain, Sai Pushpak Nandanoori, and Veronica Adetola
}
\IEEEauthorblockA{\textit{Electricity Infrastructure and Buildings Division, Pacific Northwest National Laboratory} \\ Richland, WA 99354, USA }
}

\maketitle

\begin{abstract}
Anticipated rapid growth of large digital load, driven by artificial intelligence (AI) data centers, is poised to increase uncertainty and large fluctuations in consumption, threatening the stability, reliability, and  security of the energy infrastructure. Conventional measures taken by grid planners and operators to ensure stable and reliable integration of new resources are either cost-prohibitive (e.g., transmission upgrades) or ill-equipped (e.g., generation control) to resolve the unique challenges brought on by AI Data Centers (e.g., extreme load transients). In this work, we explore the feasibility of coordinating and managing available flexibility in the grid, in terms of grid-forming storage units, to ensure stable and reliable integration of AI Data Centers without the need for costly grid upgrades. Recently developed bi-layered coordinated control strategies --- involving fast-acting, local, autonomous, control at the storage to maintain transient safety in voltage and frequency at the point-of-interconnection, and a slower, coordinated (consensus) control to restore normal operating condition in the grid --- are used in the case studies. A comparison is drawn between broadly two scenarios: a network of coordinated, smaller, distributed storage vs. larger storage installations collocated with large digital loads. IEEE 68-bus network is used for the case studies, with large digital load profiles drawn from the MIT Supercloud Dataset.

\begin{IEEEkeywords}
    AI data centers, large loads, grid-forming battery storage, coordinated controls, fault ride-through, wide-area oscillations.
\end{IEEEkeywords}

\end{abstract}

\section{Introduction}

The rapid expansion of large electric loads, such as AI data centers, cryptocurrency mining operations, and high-performance computing facilities, is introducing new operational challenges for electric power systems. Projections indicate that data centers alone could account for up to 12\% of total U.S. electricity demand by 2028, a sharp increase from just 4.4\% in 2023 \cite{council2025assessment,quint2025practical}. 
A defining characteristic of these loads is their highly dynamic power demand. Enabled by fast-acting power electronic controls, they can ramp power consumption rapidly in response to computational workloads, imposing significant stress on grid operations. Among key concerns include steep load ramps, abrupt disconnection from the grid with transfer to co-located backup generation, and the triggering of system-wide oscillations. Under certain conditions, these events may lead to frequency, angle, or voltage instability, affecting both local and system-wide reliability.

The North American Electric Reliability Corporation (NERC) Large Load Task Force (LLTF) findings emphasize the growing need for fast-responding, grid-supportive technologies that can maintain stability during such disturbances \cite{nerc2025large}. Battery energy storage systems (BESS), with their rapid response characteristics and flexible control capabilities, are increasingly recognized as a promising solution for mitigating these transient impacts and enhancing grid resilience \cite{zhu2021comprehensive,o2019use}. Recent presentations at the NERC LLTF have highlighted deployments of grid-forming (GFM) Tesla Megapacks at hyperscale data centers for AI load smoothing and low-voltage ride-through (LVRT) support. Although early deployments of large-capacity BESS co-located with load sites have shown promise in attenuating local transients, such centralized systems may not always provide the most effective or resilient system-wide support. In many cases, deploying a network of modest-sized BESS units across the grid offers improved spatial coverage, enhanced controllability under localized disturbances, and greater resilience to single-point failures. 

Recent work in this area has demonstrated \cite{o2019use,nguyen2022analysis,chatterjee2024grid,hossain2025coordinated,chatterjee2025frequency} that distributed storage networks, particularly those employing coordinated control strategies, are more effective in damping system-wide oscillatory modes, containing the spatial propagation of grid disturbances, and providing localized fast-frequency support to mitigate the severity of frequency events. These studies introduced the concept of an ``\emph{embedded storage network}", where a collection of BESS resources are strategically placed at load buses and operate as grid assets, forming a dynamic buffer between the bulk power system (BPS) and the connected load. This configuration enables effective absorption of transients originating from either side, improving system robustness under a wide range of disturbance scenarios.

Building on this concept, this paper extends the formulation to evaluate the effectiveness of distributed grid storage, embedded within the BPS, in mitigating reliability risks associated with large AI computation loads at data centers. We compare the dynamic performance of a networked distributed storage architecture against that of large-scale centralized storage co-located with data center loads, specifically, assessing their ability to mitigate stability and reliability risks resulting from: 
\begin{itemize}
    \item[(1)] sudden and significant ramping of AI loads across multiple data centers, triggering wide-area oscillatory modes;
    \item[(2)] LVRT failure leading to abrupt disconnection of a data center load or transfer to backup generation during transmission system faults, followed by grid instability.
\end{itemize}
The results from the case studies are summarized for a multi-area IEEE test system incorporating large digital loads. 

\section{Test Scenarios}\label{sec:scenarios}

\textbf{Large Digital Load (LDL)}. In order to perform the LDL case studies, we adopt a multi-area IEEE 68-bus power system network where the LDLs are added to different buses in the network. Artificial intelligence (AI) data center load profiles from the MIT Supercloud Dataset, as reported in \cite{li2024unseen}, are used as examples of played-in large load in the test scenarios. Snapshots and bus locations of the large load profiles are presented in the results section. AI data center loads are sized to peak at about 1.5\,GW at their maximum, which is about 8\% of the base load of the system. The transient dynamic modeling of AI data center loads is an active area of research \cite{jimenez2025data}. In this work, we adopt simplified ZIP (i.e., constant impedance/current/power) models for the AI data center loads, and instead focus our attention to the mitigation of the adverse impact of large loads via coordinate flexible storage units.

\textbf{Grid-Forming (GFM) Storage Network}. In order to demonstrate the effectiveness of coordinated control strategies in utilizing grid-tied storage units to mitigate the adverse impact of large AI data center loads, we consider a variety of coordinated storage network scenarios. The storage units are assumed to be interfaced with GFM inverters, as per the WECC-approved REGFM\_A1 model \cite{du2023model,kwon2024coherency}. In each case study, the cumulative power rating of the storage units is fixed at 10\% of system's base load (or, about 1.84\,GW). Broadly, two specific scenarios of storage deployment are considered:
\begin{enumerate}[label=\arabic*)]
    \item \textit{Embedded Storage Network.} Recent efforts \cite{o2019use,chatterjee2024grid,hossain2025coordinated,chatterjee2025frequency} have explored the concept of \textit{embedded storage network} which enables a network of ubiquitous grid-tied storage units, placed critically at the transmission-distribution (T\&D) substation buses to support system's resilience and operational security via coordinated, non-market, pathways. As demonstrated in \cite{hossain2025coordinated}, the embedded storage network accommodates multi-layered, autonomous and coordinated control strategies, offering multi-timescales grid support against adversarial events. Leveraging this coordinated storage network architecture, this work investigates the impact of varying penetration (from 20\%-100\%) of embedded storage units, as a fraction of the number of T\&D load buses with a storage unit.
    \item \textit{Collocated Storage.} Collocated battery energy storage systems \cite{donovan2024understanding,jones2025battery} are often considered for deployment at the AI data centers, offering benefits such as smoothening of power loads, Low Voltage Ride-Through (LVRT) support, and provision of back-up power. In this work, we consider the scenarios with collocated storage at the LDL centers, and compare those with the embedded storage network configurations.
\end{enumerate}

\textbf{Multi-Layered Coordinated Control.} Control and coordination strategies for GFM-storage units can be broadly grouped into: 1) local, autonomous, primary-layer controls (e.g., P-f and Q-V droop), and 2) secondary-layer system-wide coordination (centralized or distributed) \cite{ahmed2020review,hossain2025coordinated}. The work in \cite{hossain2025coordinated} demonstrated the benefits of a two-layered coordinated control strategy (referred to as \textit{safety-consensus}) which enforces transient safety limits on voltage and frequency (e.g., as per IEEE Standards \cite{ieee1547}) at the inverter terminals , while also providing system-level voltage and frequency regulation and power-sharing via consensus control algorithms. We adopt the safety-consensus strategy from \cite{hossain2025coordinated} as the coordinated control scheme for managing the GFM storage units in this work.

\textbf{Protection Measures}. Test setups are equipped with several frequency and voltage protection measures as per applicable NERC standards and recommendations. In particular the following are used in the case-studies presented in this work:
\begin{itemize}
    \item Low (and high) voltage ride-through (LVRT) on the LDLs, measured at the point of interconnection, as per recent ERCOT/ITIC recommendations \cite{billo2023large}.
    \item Voltage- and frequency ride-through on the GFM inverters, as per NERC Standard PRC-029-1 \cite{nerc029}.
    \item Voltage- and frequency ride-through on the synchronous generators, as per NERC Standard PRC-024-4 \cite{nerc024}.
    \item Under-frequency load-shedding (UFLS) schemes are implemented at each of the system's load buses, as per NERC Standard PRC-006-NPCC-2 \cite{nerc006}.
\end{itemize}

\section{Case Studies}\label{sec:results}

In this work, we focus our attention to two types of case-studies: one that addresses the grid instability events following a line-to-ground fault near the LDL buses, and the other on wide-area oscillations triggered by large load profiles.

\begin{figure*}
    \centering
    \includegraphics[width=0.3\linewidth]{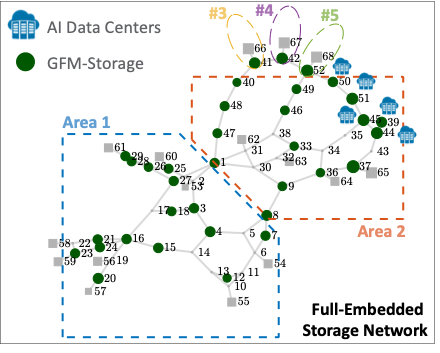}\qquad
    \includegraphics[width=0.3\linewidth]{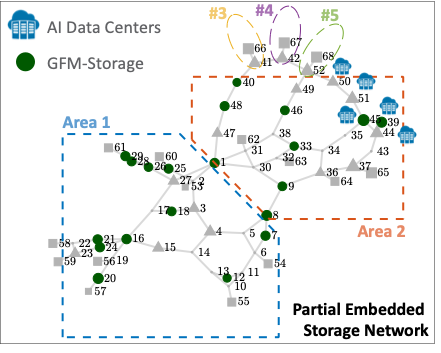}\qquad
    \includegraphics[width=0.3\linewidth]{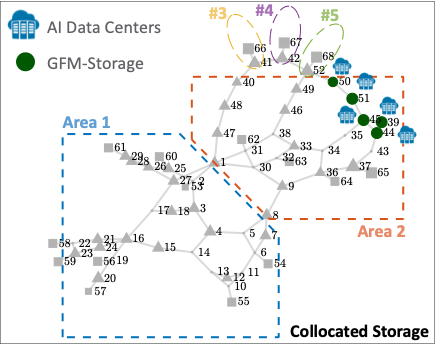}
    \caption{(Case-1) Locations of the LDL and different storage deployment scenarios on IEEE 68-bus: 1) \textit{(left)} a fully-deployed \textit{embedded storage network} with a GFM-storage at every load bus; 2) \textit{(center)} a partially deployed distributed storage network with a GFM-storage at 57\% of the load buses; and 3) \textit{(right)} collocated GFM-storage units only at the LDL buses.}\vspace{-0.1in}
    \label{fig:config}
\end{figure*}

\subsection{Case 1: Fault-Induced Grid Instability}
\begin{figure}[ht]
    \centering
    \includegraphics[width=0.8\linewidth]{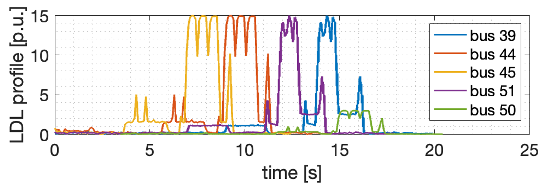}
    \caption{(Case-1) LDL profiles played-in at buses 39, 44, 45, 50, and 51. Here 1\,p.u. corresponds to 100\,MW.}\vspace{-0.1in}
    \label{fig:load_profile}
\end{figure}

Five LDLs, peaking at about 1.5\,GW (or, 15\,p.u. with 100\,MW as base apparent power), are placed at buses 39, 44, 45, 50, and 51 within area-2 of the IEEE 68-bus network, as shown in Fig.\,\ref{fig:config}. LDL profiles are drawn from the MIT Supercloud Dataset, as reported in \cite{li2024unseen}, and shown in Fig.\,\ref{fig:load_profile}. Different storage deployment configurations are considered, as shown in Fig.\,\ref{fig:config}, including, a fully deployed \textit{embedded storage network} (i.e., a GFM-storage at every load bus), a partially deployed distributed storage network with a GFM-storage at roughly 57\% of the load buses, and a scenario with GFM-storage units only at the LDL buses. In each configuration, the cumulative power rating of the storage units is 1.84\,GW. 

A line-to-ground fault is applied mid-way on the line connecting the buses 50 and 52 at time t=12.25\,s, which is cleared after 200\,ms by disconnecting the line. The line is re-connected after a delay of 375\,ms. In absence of the LDLs (and any GFM-storage unit), the grid maintains stability after fault clearance. However, as our case studies demonstrate, the fault-induced tripping of the LDLs may trigger a grid instability, especially in the absence of coordinated GFM-storage units.

\begin{figure*}
    \centering
    \hspace{-0.15in}\begin{subfigure}{0.345\textwidth}
    \centering
        \includegraphics[width=\linewidth]{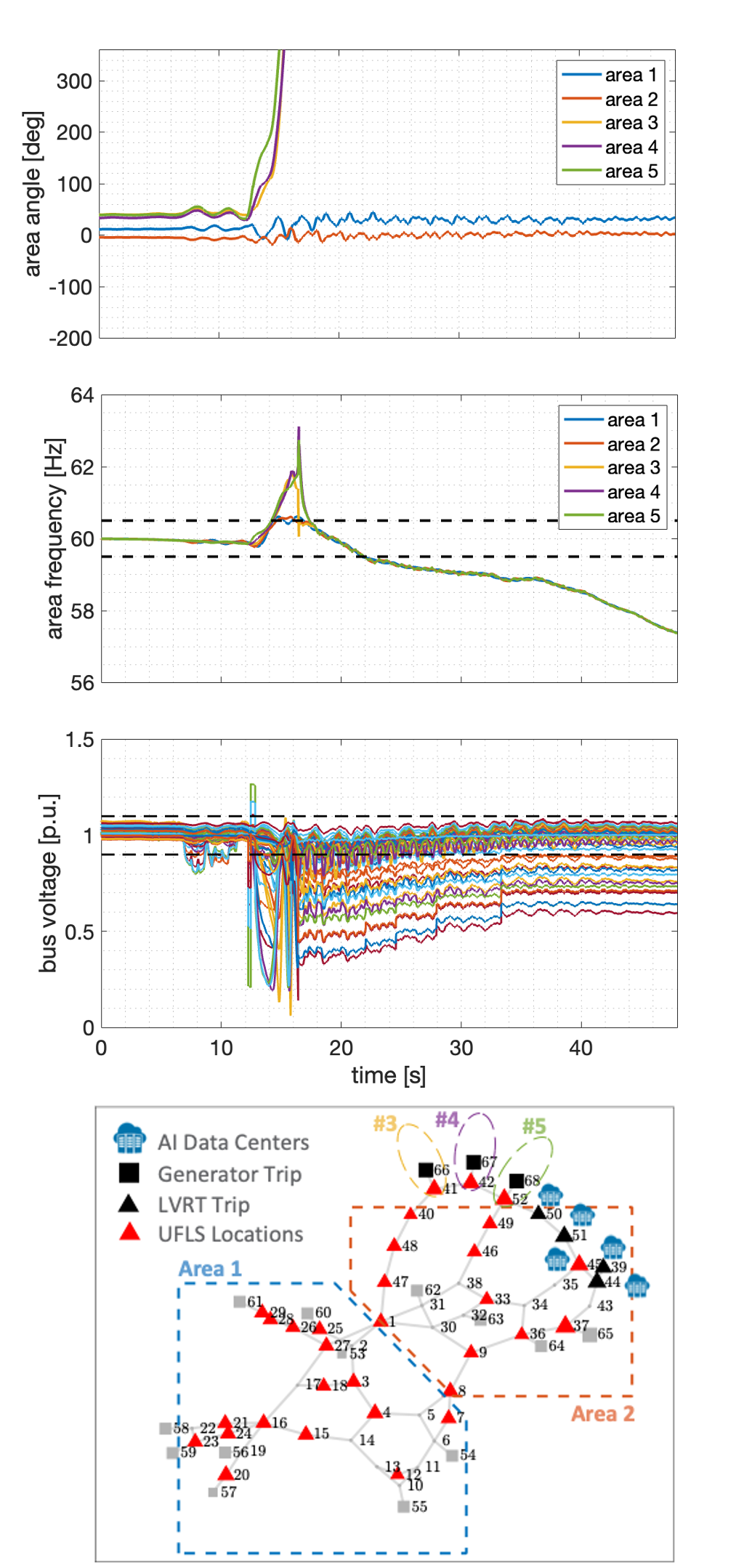}
        \subcaption{no storage}
        \label{fig:LDL}
    \end{subfigure}\hspace{-0.15in}
    \begin{subfigure}{0.345\textwidth}
    \centering
        \includegraphics[width=\linewidth]{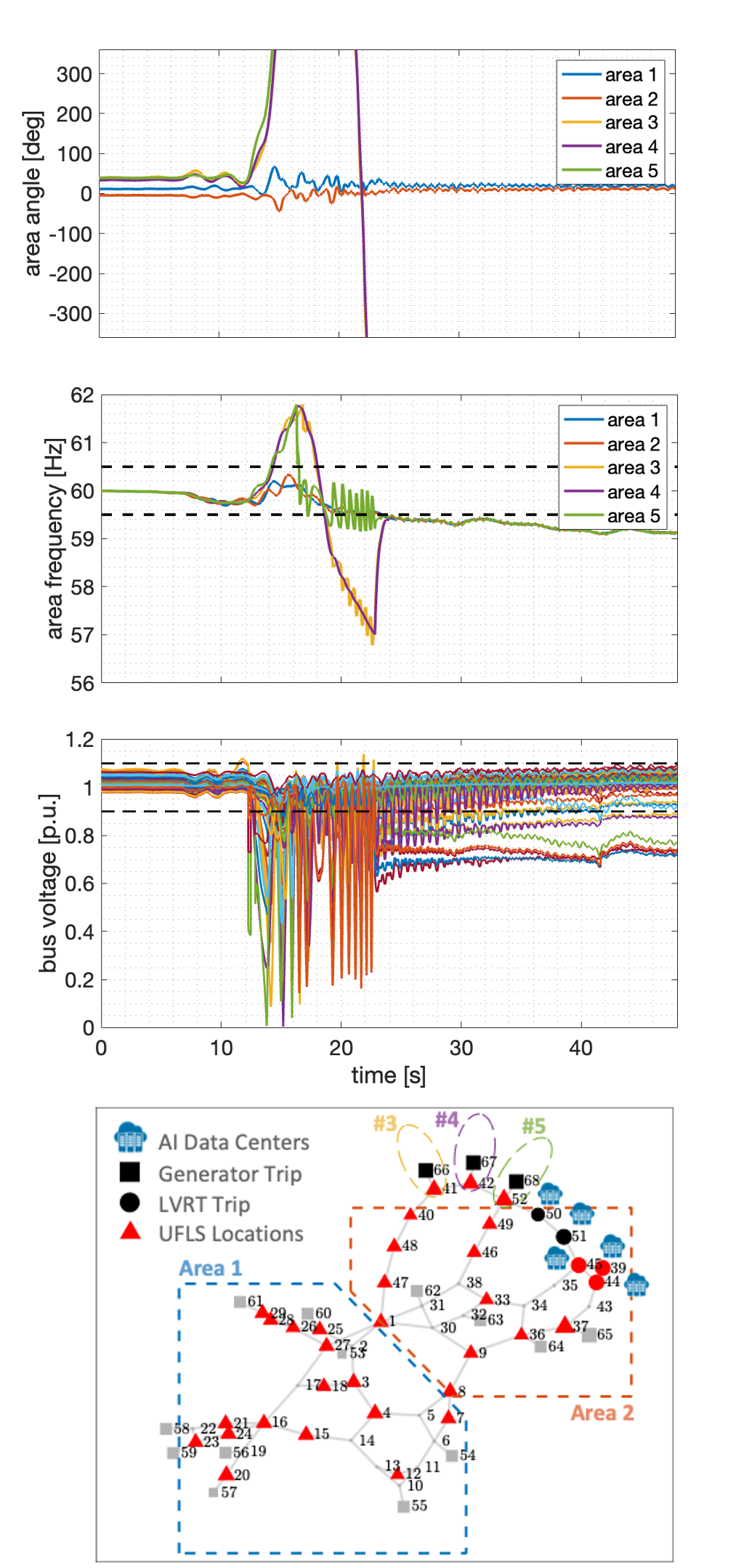}
        \subcaption{collocated GFM-storage}
        \label{fig:coloc}
    \end{subfigure}\hspace{-0.15in}
    \begin{subfigure}{0.345\textwidth}
    \centering
        \includegraphics[width=\linewidth]{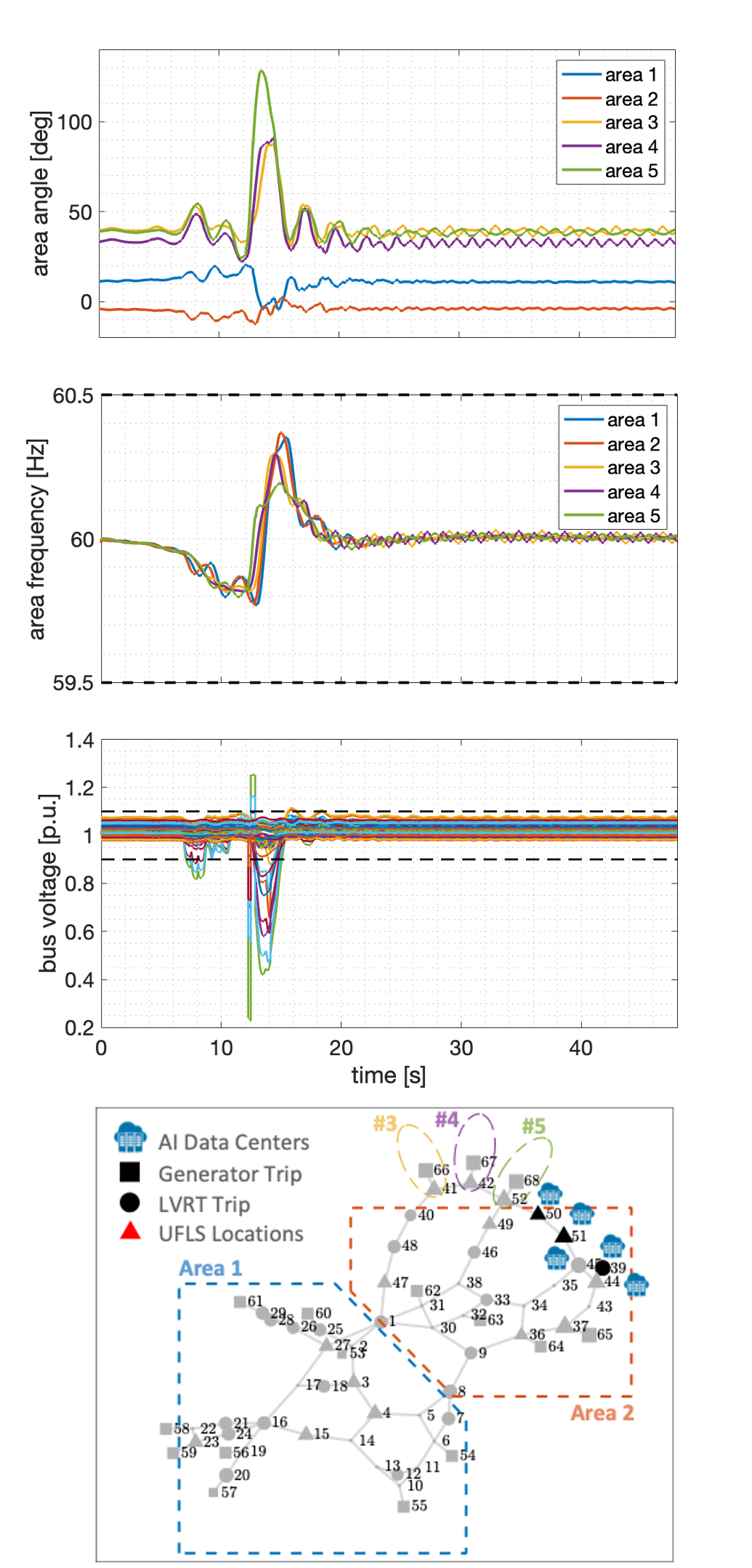}
        \subcaption{57\% load buses with GFM-storage}\label{fig:partial}
    \end{subfigure}\hspace{-0.15in}    
    \caption{Case-1 simulation results on three test scenarios with different GFM-storage deployment: a) no storage, b) collocated storage, and 3) partially (57\% load buses) deployed storage network. COI angle, COI frequency, and bus voltages are displayed, along with a plot displaying the locations with LVRT-triggered tripping of LDL, the buses undergoing UFLS, and the locations of generators tripped during the simulation. Observation: LVRT-based LDL tripping reduces in (b) compared to (c) or (a), due to the presence of large collocated storage units. But the collocated storage could not maintain grid stability.}\vspace{-0.1in}
    \label{fig:case1}
\end{figure*}

\textbf{Case 1a (Base Case): LDLs Only.} Fig.\,\ref{fig:LDL} shows the center-of-inertia (COI) angle and the COI frequency for the five areas, the bus voltages, and a graphical plot of the locations of the generators that tripped, the LDLs that tripped under LVRT, and the load buses that underwent UFLS for the scenario with only LDLs and no storage. Total four large loads tripped during fault, sequentially at t=12.40\,s, 12.45\,s, 13.26\,s, and 13.52\,s, respectively at buses 50, 51, 39, and 44. Tripping of LDLs then trigger the nearby large generators to trip at buses 67, 66, and 68 at t=15.82\,s, 15.99\,s, and 16.42\,s, respectively.

\textbf{Case 1b: Collocated Storage at LDL buses.} Fig.\,\ref{fig:coloc} shows the COI angle, COI frequency, bus voltages, and a graphical plot of the locations of the tripped generators, the tripped LDLs, and the UFLS locations for the scenario with collocated storage units. A two-layered coordinated control based on the \textit{safety-consensus} approach from \cite{hossain2025coordinated} is implemented. A clear advantage of collocated storage is evident in a reduction in the number of tripped LDLs: only two LDLs tripped at t=12.40\,s and 13.49\,s, respectively at buses 50 and 51. Tripping of LDLs then trigger the nearby large generators to trip at buses 68, 66, and 67 at t=16.30\,s, 16.79\,s, and 22.82\,s, respectively. In comparison with the base case (Case-1a), we notice two improvements: the LDLs at 39 and 44 were able to ride-through the fault without tripping, and the generators at 66, 67, and 68 tripped at later times.

\textbf{Case 1c: Partial-Embedded Storage Network.} We consider the partial embedded storage network with a GFM-storage at 57\% load buses (as shown in Fig.\,\ref{fig:config}). Fig.\,\ref{fig:partial} shows the COI angle, COI frequency, bus voltages, and a graphical plot of the locations of the tripped generators, the tripped LDLs, and the UFLS locations for the scenario with the partial-embedded storage network. The storage units are coordinated via two-layered safety-consensus approach \cite{hossain2025coordinated}. The coordinated storage network is able to stabilize the grid during and after the fault-induced transients, with no UFLS-activated load shedding and no generation loss. There are three LVRT-based tripping of LDLs at t=12.40\,s, 12.45\,s, and 14.11\,s, respectively at buses 50, 51, and 39.

\begin{table*}[ht]
    \centering
    \small
    \begin{tabular}{|c|c|c|c|r|c|c|c|r|}
         \hline 
         \multirow{2}{*}{\textbf{Scenario}} & \multicolumn{4}{c|}{\textbf{without coordination}} & \multicolumn{4}{c|}{\textbf{layered (safety-consensus) control}} \\
         \cline{2-9}
         & TSI & Gen Loss & UFLS & LDLs Tripped &  TSI & Gen Loss & UFLS & LDLs Tripped\\
         \hline\hline
         No Storage & -0.13 & 6.8\,GW & 5.4\,GW & 39, 44, 50, 51 & -- & -- & -- & -- \\
         \hline
         Collocated & -0.52 & 6.8\,GW & 5.4\,GW & 50, 51 & -0.37 & 6.8\,GW & 4.1\,GW & 50, 51\\
         \hline
         23\% Embedded & -0.34 & 6.8\,GW & 3.6\,GW & 39, 44, 50, 51 & -0.29 & 6.8\,GW & 1.4\,GW & 39, 50, 51\\
         \hline
         57\% Embedded & -0.37 & 6.8\,GW & 5.4\,GW & 39, 50, 51 & \textit{0.81} & \textit{0\,GW} & \textit{0\,GW} & 39, 50, 51\\
         \hline
         69\% Embedded & -0.35 & 6.8\,GW & 5.3\,GW & 39, 50, 51 & \textit{0.81} & \textit{0\,GW} & \textit{0\,GW} & 39, 50, 51\\
         \hline
         Full-Embedded & \textit{0.80} & \textit{0\,GW} & \textit{0\,GW} & 39, 50, 51 & \textit{0.83} & \textit{0\,GW} & \textit{0\,GW} & 39, 50, 51\\
         \hline
    \end{tabular}
    \caption{\small Case-1 summary results on grid stability performance at varying penetration of storage network, with and without layered safety-consensus control \cite{hossain2025coordinated}. \textit{Note: TSI = {Transient Stability Index}}. Collocated storage scenario fails to assure grid stability even with layered coordination, albeit preventing LVRT-triggered LDL tripping at buses 39, 44, and 45. At low (23\%) penetration of embedded storage network (with or without coordination) the grid remains unstable post-fault, although UFLS is reduced. At medium (57\%, 69\%) penetration levels of embedded storage network, layered coordination helps restore grid stability with $TSI>0$. A fully-embedded storage network assures grid stability, even without the need for coordinated controls.}\vspace{-0.1in}
    \label{tab:main}
\end{table*}

\textbf{Cases with Varying Storage Penetration [\%].} Next we evaluate the system performance, with and without the layered coordination scheme, under varying distributed storage penetrations. Table\,\ref{tab:main} summarizes the results. Four performance measures are used to compare the results across scenarios:
\begin{itemize}
    \item Transient Stability Index (TSI), measured as:
    \begin{align*}
        \vspace{-0.1in}\text{TSI} &= {\left(360^o - \delta_{\max}\right)}/{\left(360^o + \delta_{\max}\right)}\vspace{-0.1in}
    \end{align*}
    where $\delta_{\max}$ (in degree) is the maximum angular excursion observed anywhere in the system during simulations.
    \item Aggregated loss of generating resources (Gen Loss) observed during simulations.
    \item Aggregated UFLS-triggered load loss during simulations.
    \item Total number of LDLs that tripped due to LVRT.
\end{itemize}
We observe that at low (23\%) penetration of embedded storage network (with or without coordination) the grid remains unstable post-fault, although UFLS is reduced from the base case. At medium (57\%, 69\%) penetration levels of embedded storage network, layered coordination helps restore grid stability with $TSI>0$, no generation loss, and no UFLS-triggered load loss. Finally, a fully-embedded storage network assures grid stability, even without the need for coordinated controls.

\subsection{Case 2: Wide-Area Oscillations}

\begin{figure}
    \centering
    \hspace{-0.2in}\includegraphics[width=0.5\linewidth]{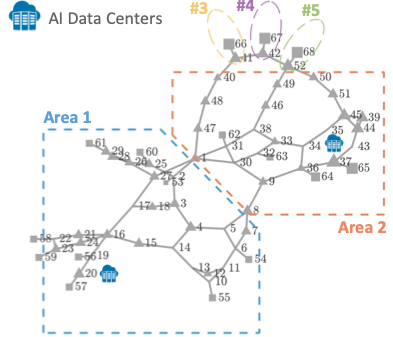}\hspace{0in}
    \includegraphics[width=0.5\linewidth]{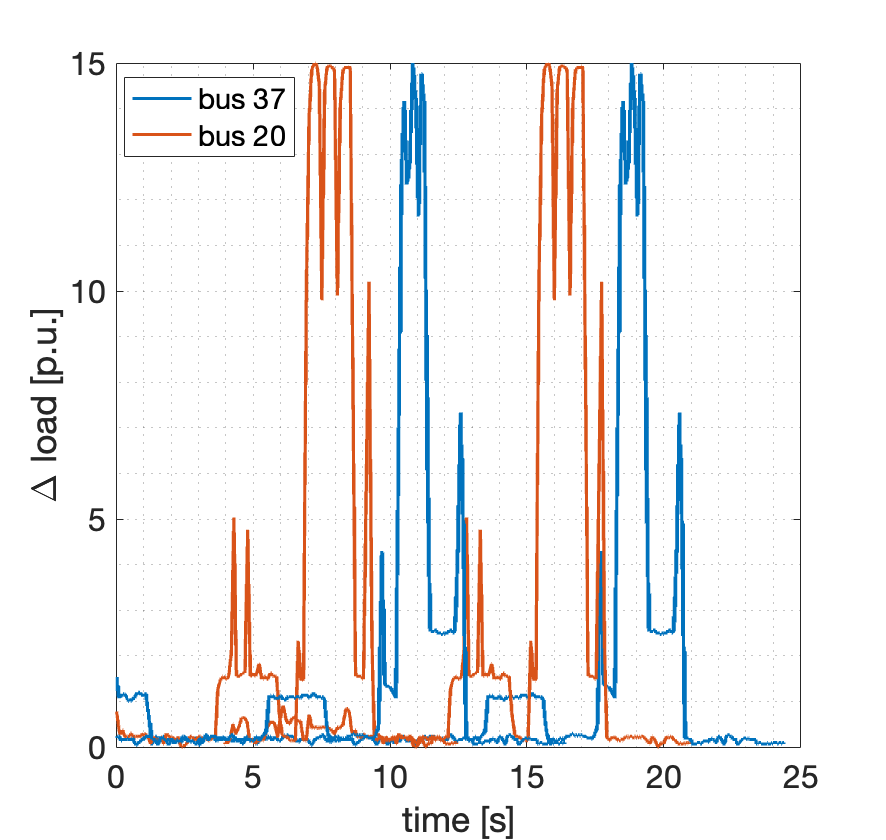}\hspace{-0.2in}
    \caption{Case-2 LDL locations and load profiles.}\vspace{-0.1in}
    \label{fig:load_profile_osc}
\end{figure}

\begin{figure}
    \centering
    \includegraphics[width=0.9\linewidth]{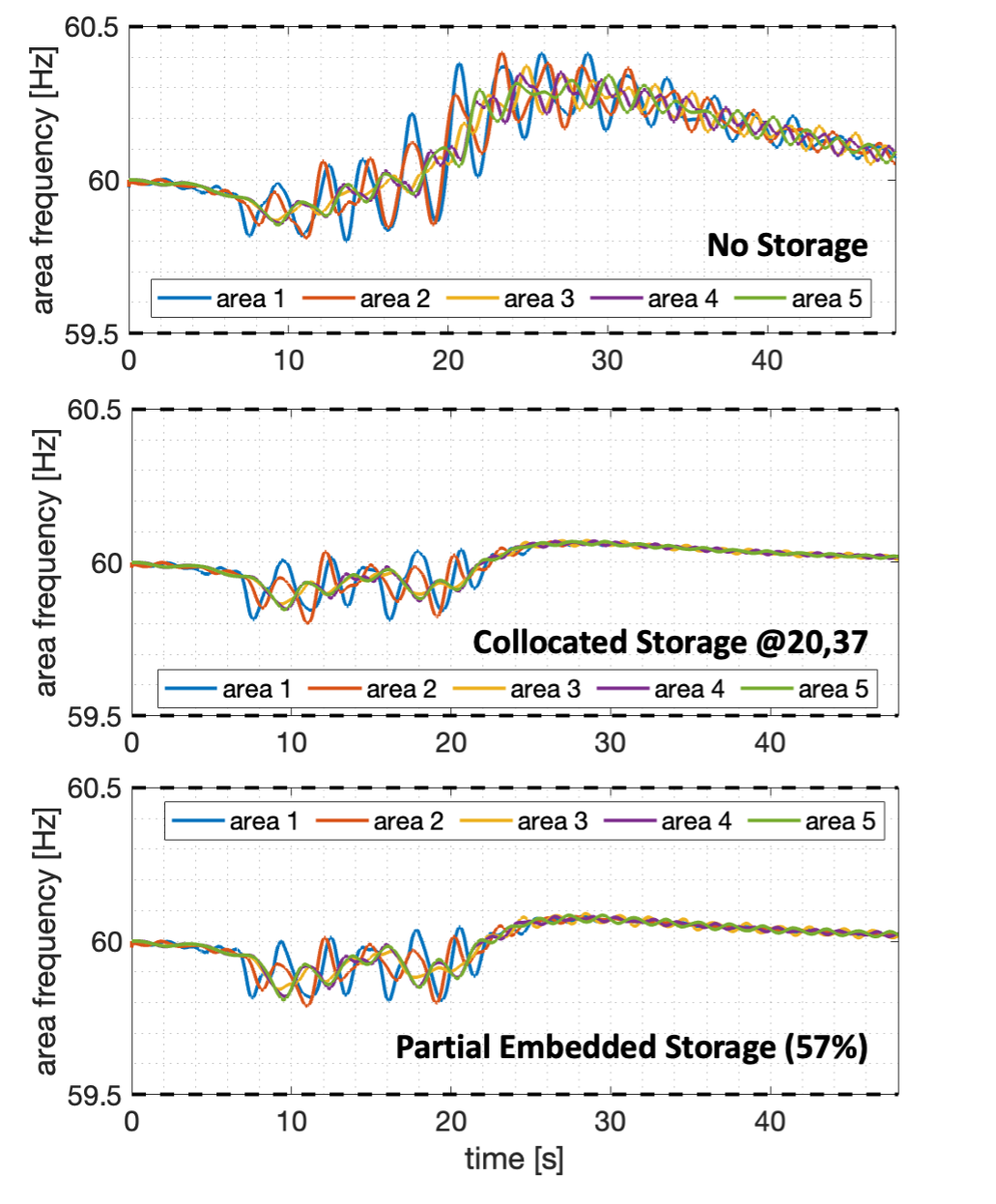}
    \caption{Case-2 simulation results on three test scenarios: the base case with no storage, a scenario with collocated storage at the LDL buses (20, 37), and the partial-embedded storage configuration from Fig.\,\ref{fig:config}. Both storage configurations dampen the wide-area oscillations triggered by LDLs.}\vspace{-0.1in}
    \label{fig:osc_results}
\end{figure}

For the wide-area oscillation studies, we place two LDLs, peaking at about 1.5\,GW, at bus 15 in area-1 and bus 37 in area-2, as shown in Fig.\,\ref{fig:load_profile_osc}. LDLs trigger wide-area oscillations, shown in Fig.\,\ref{fig:osc_results}, at the dominant inter-area modes. Two storage configurations are tested: collocated storage configuration (with GFM-storage at buses 20 and 37), and a partial-embedded storage network with a GFM-storage at 57\% of the load buses (shown in Fig.\,\ref{fig:config}).  In each configuration, the cumulative power rating of the storage units is 1.84\,GW. Bi-layered safety-consensus controls \cite{hossain2025coordinated} are used to coordinate the storage units. The results in Fig.\,\ref{fig:osc_results} demonstrate the effectiveness of either storage configurations in successfully dampening the dominant inter-area oscillations triggered by the LDLs.

\section{Conclusion}
In this work, we investigate the feasibility of mitigating the grid stability and reliability risks posed by the integration of large digital loads (e.g., AI data center loads). In particular, we study the effectiveness of coordinated multi-layered control strategies in managing the flexibility offered by grid-forming distributed storage units across the grid. Using IEEE 68-bus network as the test system, with large digital load profiles synthesized from the MIT Supercloud Dataset \cite{li2024unseen}, we demonstrate the effectiveness of bi-layered \textit{safety-consensus} coordinated control strategy \cite{hossain2025coordinated} in improving the grid stability, when sufficiently many grid-tied distributed storage units are available. While large collocated storage units improve local voltage behavior at the points-of-interconnection (thereby reducing the LVRT-triggered tripping, the overall grid stability is significantly improved when multiple, smaller, distributed storage units are coordinated via safety-consensus control. 

Future studies will generalize this work to understand the impact of speeding up large load interconnection risk evaluation, optimal placement and sizing of the embedded storage units, as well as other flexible grid assets (e.g., flexible load) as mitigating resources.

\section*{Ackowledgement}
The authors would like to thank Dr. Ian Beil of Portland General Electric and Brett Ross of Pacific Northwest National Laboratory for the fruitful discussions related to this work.

\bibliographystyle{IEEEtran}
\bibliography{main}

\begin{thebibliography}{10}
\providecommand{\url}[1]{#1}
\csname url@samestyle\endcsname
\providecommand{\newblock}{\relax}
\providecommand{\bibinfo}[2]{#2}
\providecommand{\BIBentrySTDinterwordspacing}{\spaceskip=0pt\relax}
\providecommand{\BIBentryALTinterwordstretchfactor}{4}
\providecommand{\BIBentryALTinterwordspacing}{\spaceskip=\fontdimen2\font plus
\BIBentryALTinterwordstretchfactor\fontdimen3\font minus
  \fontdimen4\font\relax}
\providecommand{\BIBforeignlanguage}[2]{{%
\expandafter\ifx\csname l@#1\endcsname\relax
\typeout{** WARNING: IEEEtran.bst: No hyphenation pattern has been}%
\typeout{** loaded for the language `#1'. Using the pattern for}%
\typeout{** the default language instead.}%
\else
\language=\csname l@#1\endcsname
\fi
#2}}
\providecommand{\BIBdecl}{\relax}
\BIBdecl

\bibitem{council2025assessment}
R.~Quint, J.~J. Zhao, and K.~Thomas, ``An assessment of large load
  interconnection risks in the western interconnection,'' Council, Western
  Electricity Coordinating, Tech. Rep., 2025.

\bibitem{quint2025practical}
R.~Quint, K.~Thomas, J.~Zhao, A.~Isaacs, and C.~Baker, ``Practical guidance and
  considerations for large load interconnections,'' Elevate Energy Consultancy,
  Tech. Rep., 2025.

\bibitem{nerc2025large}
L.~L. T.~F. (LLTF), ``{Characteristics and Risks of Emerging Large Loads},''
  North American Reliability Corporation (NERC), Tech. Rep., 2025.

\bibitem{zhu2021comprehensive}
L.~Zhu, W.~Yu, Z.~Jiang, C.~Zhang, Y.~Zhao, J.~Dong, W.~Wang, Y.~Liu,
  E.~Farantatos, D.~Ramasubramanian \emph{et~al.}, ``A comprehensive method to
  mitigate forced oscillations in large interconnected power grids,''
  \emph{IEEE Access}, vol.~9, pp. 22\,503--22\,515, 2021.

\bibitem{o2019use}
R.~S. O'Neil, A.~S. Becker-Dippmann, and J.~D. Taft, ``The use of embedded
  electric grid storage for resilience, operational flexibility, and
  cyber-security,'' Pacific Northwest National Lab.(PNNL), Richland, WA (United
  States), Tech. Rep., 2019.

\bibitem{nguyen2022analysis}
Q.~Nguyen, A.~Singhal, R.~O'Neil, J.~Taft, J.~Twitchell, and A.~Somani,
  ``Analysis of distributed energy storage as a core grid infrastructure via
  production cost modeling,'' in \emph{2022 IEEE Power \& Energy Society
  General Meeting (PESGM)}.\hskip 1em plus 0.5em minus 0.4em\relax IEEE, 2022,
  pp. 1--5.

\bibitem{chatterjee2024grid}
K.~Chatterjee, R.~R. Hossain, S.~P. Nandanoori, S.~Kundu, S.~Sinha, D.~Baldwin,
  and R.~Melton, ``Grid-forming storage networks: Analytical characterization
  of damping and design insights,'' in \emph{IEEE 63rd Conference on Decision
  and Control (CDC)}, 2024, pp. 5346--5352.

\bibitem{hossain2025coordinated}
R.~R. Hossain, K.~Chatterjee, S.~P. Nandanoori, S.~Kundu, L.~Marinovici,
  K.~Kalsi, and D.~Baldwin, ``Coordinated frequency regulation in grid-forming
  storage network via safety-consensus,'' in \emph{IEEE Electrical Energy
  Storage Applications \& Technologies Conference (EESAT)}, 2025.

\bibitem{chatterjee2025frequency}
K.~Chatterjee, R.~R. Hossain, S.~P. Nandanoori, S.~Kundu, D.~Baldwin, and
  R.~Melton, ``Frequency control and disturbance containment using grid-forming
  embedded storage networks,'' in \emph{IEEE Electrical Energy Storage
  Applications \& Technologies Conference (EESAT)}, 2025.

\bibitem{li2024unseen}
Y.~Li, M.~Mughees, Y.~Chen, and Y.~R. Li, ``{The unseen AI disruptions for
  power grids: LLM-induced transients},'' \emph{arXiv preprint
  arXiv:2409.11416}, 2024.

\bibitem{jimenez2025data}
A.~Jimenez-Ruiz and F.~Milano, ``Data center model for transient stability
  analysis of power systems,'' \emph{arXiv preprint arXiv:2505.16575}, 2025.

\bibitem{du2023model}
W.~Du~\textit{et al}, ``{Model specification of droop-controlled, grid-forming
  inverters (REGFM\_A1)},'' Pacific Northwest Nat. Lab, Tech. Rep., 2023.

\bibitem{kwon2024coherency}
K.-B. Kwon, R.~R. Hossain, S.~Mukherjee, K.~Chatterjee, S.~Kundu, S.~Nekkalapu,
  and M.~Elizondo, ``Coherency-aware learning control of inverter-dominated
  grids: A distributed risk-constrained approach,'' \emph{IEEE Control Systems
  Letters}, vol.~8, pp. 2120--2125, 2024.

\bibitem{donovan2024understanding}
P.~Donovan, ``{Understanding BESS: Battery Energy Storage Systems for Data
  Centers},'' Energy Management Research Center, Schneider Electric, White
  Paper 185, Tech. Rep., 2024.

\bibitem{jones2025battery}
S.~Jones and S.~G. Vennelaganti, ``{Battery Energy Storage Applications at Data
  Centers – Tesla’s Perspective},''
  [Online]~\url{https://www.esig.energy/event/lltf-bess-applications-tesla/},
  2025.

\bibitem{ahmed2020review}
K.~Ahmed, M.~Seyedmahmoudian, S.~Mekhilef, N.~Mubarak, and A.~Stojcevski, ``A
  review on primary and secondary controls of inverter-interfaced microgrid,''
  \emph{Journal of Modern Power Systems and Clean Energy}, vol.~9, no.~5, pp.
  969--985, 2020.

\bibitem{ieee1547}
IEEE, ``{IEEE Standard for Interconnection and Interoperability of Distributed
  Energy Resources with Associated Electric Power Systems Interfaces},''
  \emph{IEEE Std 1547-2018}, pp. 1--138, 2018.

\bibitem{billo2023large}
J.~Billo, ``{LFLTF: Large Load Voltage Ride-Through Requirements},'' ERCOT
  Operations Planning, Sep 2023.

\bibitem{nerc029}
NERC, ``{PRC-029-1 – Frequency and Voltage Ride-through Requirements for
  Inverter-based Resources},'' Oct 2024.

\bibitem{nerc024}
------, ``{PRC-024-4 – Frequency and Voltage Protection Settings for
  Synchronous Generators and Synchronous Condensers},'' Aug 2024.

\bibitem{nerc006}
------, ``{PRC-006-NPCC-2 – Automatic Underfrequency Load Shedding},'' Feb
  2020.

\end{thebibliography}

\end{document}